\documentstyle[floats,aps]{revtex}
\input psfig
\begin{document}

\title{Canonical quantum gravity with new variables and loops: a
report}

\author{Jorge Pullin\\Center for Gravitational Physics and Geometry\\
Department of Physics, 104 Davey Lab\\The Pennsylvania State
University\\
University Park, PA 16802}

\maketitle

\begin{abstract}
This is a brief and updated summary of a talk given at the
International Conference on Gravitation and Cosmology that took place
in Poona in December 1995. It is very brief and is mostly intended as
a guide to current literature, or to keep people updated only in very
broad terms on the latest developments in the subject.
\end{abstract}
\vspace{-6.5cm} 
\begin{flushright}
\baselineskip=15pt
CGPG-96/6-5  \\
gr-qc/9606061\\
\end{flushright}
\vspace{4.5cm}

\section{Introduction}

The relationship of quantum mechanics and gravity has been a
problematic one since the early attempts to quantize the theory of
general relativity. There are a plethora of {\em prima facie}
difficulties (see \cite{Is} for a summary) that one faces even before
laying down any specific approach to the subject. This has discouraged
many people away from the field. On the other hand, it is not healthy
to just not attack a problem because some obvious difficulty is
forecast to appear based on general considerations. It is much more
reasonable to devote some effort to a detailed analysis, since in many
cases the particular details of how the expected difficulty appears
can lead to insights towards its cure.

Partly with this philosophy as a motivation, a group of people have
been pursuing several aspects of the canonical quantization of
gravity. The canonical approach faces some a priori problems of its
own, like how to recover a spacetime picture from three dimensional
notions quantum mechanically, or the related issue of the ``problem of
time'' (see \cite{Ku} for a review). Yet, it also seems as an
appropriate setting for elucidating issues like what is exactly the
space of states of the theory or how degrees of freedom and symmetries
interlace in the theory.  Some of these latter issues were also
considered a stumbling block for the theory early on: how to
characterize the space of states given that spatial diffeomorphism
invariance was a symmetry and how to solve the nonpolynomial
constraints of the theory quantum mechanically were one of the
deterrents of early progress of the subject in the 60's (see for
instance \cite{DeWi}).

Unlike the difficulties mentioned at the beginning (problem of time,
observables) some progress on the issue of diffeomorphism invariance,
space of states  and solutions to the constraints was achieved through
the introduction of a new set of variables that allows to describe
general relativity in terms of notions closer to Yang--Mills theory. 
These new variables were introduced by Ashtekar \cite{As} about ten
years ago and led to a large amount of new insights and perspectives on
canonical quantum gravity. In the canonical approach one describes
spacetime through the initial data on a spatial slice consisting of a
three-dimensional metric $q^{ab}$ and the extrinsic curvature $K_{ab}$.
The new variables replace these by a set of (densitized) triads
$\tilde{E}^a_i$ and a canonically conjugate momentum that transforms
under space and triad transformations as an $SO(3)$ connection $A_a^i$. 
The spacetime diffeomorphism invariance of general relativity translates
itself into four constraints that the variables have to satisfy, three
of them representing spatial diffeomorphism. The formalism is also
invariant under triad rotations and this translates itself into three
additional constraints that have exactly the form of an $SO(3)$ Gauss
law. Therefore the reduced phase space of the theory is exactly that of
an $SO(3)$ Yang--Mills theory with four additional constraints. 

An additional element that enters into play is the fact that the
connection $A_a^i$ is a complex quantity. If one constructs the
connection and triad from a slice of a real four dimensional metric, the
formalism assures that the evolution keeps everything real forever. One
is simply using a complex coordinatization on a real phase space.
However, in the quantum theory one has to ensure that the resulting
metric and its time derivatives be real. This can be implemented through
a set of ``reality conditions'', a set of constraints that turns out to
be second class. Another approach is to forget these constraints and
then choose in the quantum theory an inner product that makes the
observable quantities self-adjoint. This, in fact, may help to select
the correct inner product for the theory, something the Dirac
quantization procedure says nothing about \cite{Re93}.

An approach that has proved fruitful for the gauge invariant
description of Yang--Mills theories is the use of holonomic variables,
in which one codes the gauge invariant information of the theory in
the holonomies along families of closed curves. One can actually build
an entire quantum representation in which the wavefunctions are purely
functionals of loops, called the ``loop representation''. This
approach was pioneered in Yang--Mills theories by Gambini and Trias
\cite{GaTr81,GaTr86} and was applied in gravity by Rovelli and Smolin
\cite{RoSm88}. When applied to gravity, it immediately led to remarkable
insights. On the one hand the diffeomorphism invariance of the theory
gets naturally embodied into diffeomorphism invariance of closed
curves and therefore the quantum states of the theory are {\em knot
invariants}. Moreover the Hamiltonian constraint of general
relativity, which embodies all the dynamical information of the
theory, gets coded into a rather simple operator that acts at
intersections of the knots \cite{RoSm88,Ga91,BrPu93,Bl}.

All these ideas were developed early on. Over the last two years there
has been considerable progress in the consolidation of many of these
ideas, taking them from being ``handwaving'' arguments to either
rigorous theorems or rather extensive formulations for doing practical
computations given certain assumptions.  In particular new connections
with knot and graph theory have been drawn. In this talk I will
briefly summarize three directions in which the formalism is
advancing: a) the implementation of a ``generalized Wick transform" to
deal with the issue of the reality conditions; b) the development of
spin networks as a basis of states in the loop representation; c) The
use of a lattice regularization for the Hamiltonian constraint that
suggests unexpected dynamical connection between gravity and knot
theory. This paper is only intended as a pointer to current
literature, to which we refer the reader for additional details.

\section{Generalized Wick transform}

As we mentioned in the introduction, the new variables are complex
variables. Specifically, the connection $A_a^i = \Gamma_a^i +i K_a^i$
where $\Gamma_a^i$ is the spin connection built from the triad and
$K_a^i$ is the extrinsic curvature with one index raised with the
triad.  As suggested before, one possible approach is to treat quantum
mechanically $A$ and $\tilde{E}$ as complex independent variables and
then choose an inner product that ensures that one has a real
spacetime
metric. The details of how this is achieved depend on at which level
one wants to introduce an inner product. If one introduces an inner
product before imposing the constraint, then one can simply require
the metric and its Poisson bracket with the Hamiltonian to be
real. However, a rather general consensus is that one only introduces
a physical inner product in canonical quantization after imposing the
constraints. Since neither the metric nor its Poisson bracket with the
Hamiltonian commute with the constraints, one cannot make statements
about them after imposing the constraints. What one has to do is to
find conditions on the observables of the theory (quantities that
commute with the constraints) that are equivalent to the reality of
the four dimensional metric. Unfortunately, a practical impediment is
that we do not know explicitly any quantities that commute with the
constraints of general relativity that may be used for this purpose,
and they are expected to be nonlocal expressions \cite{AnTo,GoLeSt}.
This has led some people to be suspicious of the whole approach since
one is ``sweeping under the rug'' of the impossibility of finding
observables the problem of implementing the reality conditions. This
feeling is reinforced by the fact that the complexity of the variables
is crucial: one can build real connection-based formulations but they
have high order nonpolynomial constraints \cite{Ba}. It therefore
seems one is concealing in these unsolved difficulties an important
issue.

A lot of work on the subject has simply ignored the issue of the reality
conditions. At a certain level, it makes sense. Suppose one is looking
for solutions to the constraints of the theory, or of ways of
characterizing the space of solutions. It is natural to start the search
ignoring completely the inner product and then, from the solutions
found, pick the ones of physical interest by requiring that they be
normalizable with respect to the inner product. At that level one would
then worry about having an inner product that respects the reality
conditions. The big worry is: will the reality conditions force upon us
an inner product such that most of the work done solving the constraints
is rendered useless? 

An important step away from the above worry was a recent construction
introduced (in slightly different flavors) independently by Thiemann
\cite{Th} and Ashtekar \cite{As96}. What follows is an account of their
work. The idea is actually simple and
attractive. Suppose one starts with general relativity formulated
canonically in the traditional way, but using triads instead of the
metric as fundamental variables. One then introduces a canonical
transformation to a new set of variables given by the triad and a real
connection $A_a^i = \Gamma_a^i +K_a^i$ (we use the same notation that
for the complex connection used throughout the rest of the paper). The
theory has a Gauss law and diffeomorphism constraints that look in
terms of these variables exactly the same as those of the complex
Ashtekar formulation. However, if one writes the Hamiltonian
constraint of ordinary general relativity with a Lorentzian signature
in spacetime, it turns out to be a nonpolynomial expression \cite{Ba}.
This is why the Ashtekar variables had to be complex, the complexity
achieving the simplification of the Hamiltonian constraint to a
polynomial expression. If one were interested in general relativity
with a Euclidean signature, the Hamiltonian constraint in terms of
the real variables we just introduced is exactly the same as that of
the usual Ashtekar formulation, $H=\epsilon_{ijk}
\tilde{E}^a_i\tilde{E}^b_j F_{ab}^k$. This actually has been known for
quite some time \cite{To}.

The new insight consists in noticing that one can construct a canonical
transformation  that maps the constraints of the Euclidean theory to
those of the Lorentzian theory. One can then simply work in the
Euclidean theory (which is tantamount to working in the usual Ashtekar
formulation but assuming that everything is real) and one knows that the
resulting theory can be canonically mapped to real Lorentzian general
relativity. In a sense, this procedure ``legitimizes'' many calculations
that have been done ignoring the reality conditions and assuming the
Ashtekar variables were real. The generator of the canonical
transformation is actually very simple,
\begin{equation}
T = {i \pi \over 2} \int d^3x K_a^i \tilde{E}^a_i.
\end{equation}

The main drawback of the construction is that if one wishes to
implement the canonical transformation quantum mechanically, the
operator that materializes it is given by the exponential of $T$,
which yields a highly complicated factor ordering for the resulting
constraints. It is yet to be investigated what are the implications of
this fact for the issue of the constraint algebra and other details of
the canonical quantization, like the Hilbert space of solutions of
constraints. On the positive side, the construction seems to work well
not only for vacuum relativity but also for the theory coupled to
matter. An interesting point to notice is that the proposed method
maps solutions of the constraints of one theory to the other. It does
not, however, map four dimensional Riemannian solutions to Lorentzian
ones. However, it maps the integral curves of the Hamiltonian on the
constraint surface from one theory to the other.  This might lead to
new dynamical insights in classical gravity. The main achievement however,
is that previous calculations that were only heuristic, because they
either ignored the reality conditions or outright treated the variables
as real, can now find applicability in the Lorentzian domain, possibly 
in a rigorous way.

\section{Spin networks}

The idea of a loop representation is to encode all the gauge invariant
information of the theory into holonomies along loops.  This is easily
understood through the loop transform, which is one of the possible ways
of defining the loop representation,
\begin{equation}
\psi(\gamma) = \int D A\, W_\gamma[A] \psi[A]
\end{equation}
where $\psi[A]$ is a wavefunction in the connection representation,
$W_\gamma[A]$ is the trace of the holonomy (Wilson loop) of the
connection $A$ along the loop $\gamma$, and $DA$ is a measure in the infinite
dimensional set of all connections.

This transform is analogous to a Fourier transform. One is expanding the
functional $\psi[A]$ in a ``basis'' of functionals parameterized by a
continuous parameter $\gamma$, $W_\gamma[A]$. Apart from the many
mathematical subtleties involved in the fact that this is really a
functional transform, one important difference stands out: the ``basis"
formed by $W_\gamma[A]$ is an overcomplete one. There are identities
satisfied by Wilson loops, called Mandelstam identities. For instance,
for the case of $SU(2)$, one has that for two loops $\alpha$ and
$\beta$, as shown in figure 1,
\begin{equation}
W_\alpha[A] W_\beta[A] = W_{\alpha\circ\beta}[A] +
W_{\alpha\circ\beta^{-1}}[A]
\end{equation}
and also $W_\gamma[A]=W_{\gamma^{-1}}[A]$ and $W_{\alpha\circ\beta}[A]
= W_{\beta\circ\alpha}[A]$, where $\circ$ denotes composition of
curves at an origin, which for simplicity can be taken at the
intersection of $\alpha$ and $\beta$. These identities are nonlinear
and through combinations of them one can be led to many
nontrivial-looking identities among Wilson loops. The identities are
inherited by the wavefunctions $\psi(\gamma)$ in the loop
representation.

The presence of these identities makes the space of functions of loops
$\psi(\gamma)$ quite nontrivial. As an example, for various reasons
one might be interested in wavefunctions that are one on smooth loops
and zero on loops with intersections or kinks. It turns out that such
functions would easily satisfy the Hamiltonian constraint.
Unfortunately, they fail to satisfy the Mandelstam constraints. Also
consider the following: the diffeomorphism constraint implies that
wavefunction must be knot invariants. On the other hand, very few of
the available knot invariants in the mathematical literature satisfy
identities like Mandelstam's. Actually this is not entirely true: the
Mandelstam identities require that the wavefunctions have defined
values on loops with intersections whereas most of the invariants of
the mathematical literature are only defined for smooth loops. So one
could conceive defining the values of these wavefunctions in such a
way as to make them compatible with the Mandelstam identities. This
procedure is, however, severely constrained \cite{GaPu96a}.

\begin{figure}
\psfig{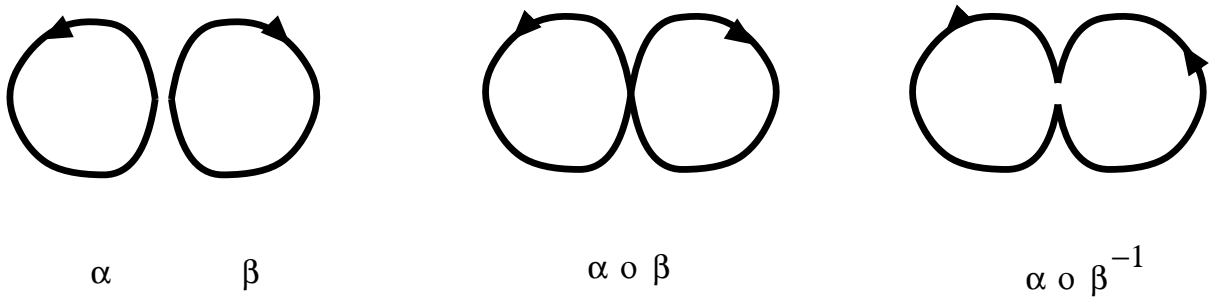}
\caption{The loops that intervene in the Mandelstam identity. In all cases
the curves are the same, we added little separations at the intersection
for a clearer visual appearance of the connectivity.}
\end{figure}
It therefore appears that it would be useful to encounter a subset of
Wilson loops that would not be related through Mandelstam identities
and would yet be sufficient to expand all gauge invariant
functions. Rovelli and Smolin \cite{RoSm95a} noticed that spin networks
provide a natural way to tabulate such a basis. In the original
version, due to Penrose \cite{Pe}, spin networks are colored trivalent
graphs in two dimensions. Colored means that to each strand connecting
two vertices a number is assigned. What does this have to do with
quantum gravity? A simple example can clarify this. Consider the loops
depicted in figure 1, which intervene in the Mandelstam identity
considered before. Now consider a diagrammatic notation as the one
introduced in figure 2. It is clear that $W(\Gamma_1)$ and
$W(\Gamma_2)$ are independent combinations of Wilson loops. One can
therefore view $\Gamma_1$ and $\Gamma_2$ as spin networks and as
generators of independent Wilson loops through the combinations
defined.  To make a long story short, this happens for all possible
loops and networks. That is, given a graph in two dimensions with
trivalent intersections, one can construct in a univocous way a basis
of combinations of Wilson loops, that are not related through
Mandelstam identities. It turns out that the reason for this is that
group-theoretic considerations of recoupling theory are at the
foundation of the spin network approach \cite{KaLi}. But for this
review we will leave matters here. It suffices to say that one can now
construct a representation in which wavefunctions are labelled by spin
networks and one can do there whatever one wished to do in terms of
loops, like define a Hamiltonian constraint \cite{Bo}, dynamics,
define a time and true Hamiltonian \cite{Ro95}. Moreover, it has been
shown rigorously using measures in infinite dimensional spaces that
the spin network states constructed as discussed above are orthogonal
\cite{Ba95}. In addition, certain operators measuring the area of a
surface and volume of a portion of space are naturally diagonalized by
the spin network basis \cite{RoSm95b}. This has profound physical
implications, since it means that areas are quantized in quantum
gravity. This may, for instance, lead to a new understanding of the
thermodynamics of black holes, since there now is a natural discrete
structure associated with the horizon and therefore one can count
states and define notions of entropy for it \cite{Ro96}.
\begin{figure}
\hskip 3cm \psfig{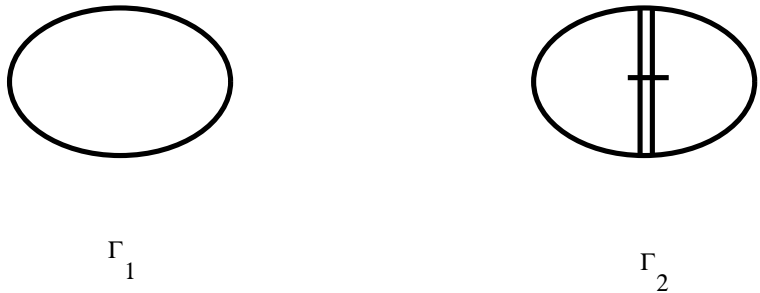}
\caption{A symbolical notation for the loops intervening in the Mandelstam
identity. $\Gamma_1$ corresponds to $\alpha\circ\beta^{-1}$ in figure 1
(if the strands that go to the intersection were parallel, one could genuinely
omit them, that is the basis for the notation). $\Gamma_2$ corresponds to
the symmetrized combination (at a Wilson loop level): $W_{\Gamma_2}[A]\equiv
W_\alpha[A]W_\beta[A]+W_{\alpha\circ\beta}[A]$}
\end{figure}

\section{Lattice regularization and the dynamics of gravity as skein
relations}

One of the central issues in the construction of a quantum theory of
gravity is the definition of a regularized quantum Hamiltonian
constraint. There are several ab-initio difficulties one can imagine
in implementing such an operator. On one hand, one has the problem
that the Hamiltonian is not diffeomorphism invariant and yet the
theory has a diffeomorphism constraint. Moreover, the Hamiltonian
involves products of basic operators, and most regularization
procedures are not diffeomorphism invariant. One can try however, to
gain some feeling for the structure of the space of
solutions of the theory without facing the entire list of difficulties
associated with regularizing the Hamiltonian. There are several
heuristic proposals for the regularization of the action of the
Hamiltonian constraint in loop space
\cite{RoSm88,Ga91,BrPu93,Bl}. Most of these proposals end up with an
operator that acts nontrivially at intersections of loops. There, the
action of the operator can be split into two pieces: a diffeomorphism
invariant action, consisting of a rerouting of the loops at the
intersection and a diffeomorphism- and regularization-dependent
prefactor. The topological piece of the action is not unique, it
originates in the need to represent the curvature that arises in the
Hamiltonian as a deformation of the loops and the action of the the
triads as a rerouting and there are differing proposals on how to accomplish
this. The prefactor absorbs the distributional
nature of the functional derivatives that arise in the definition of
the quantum triad operators.  The implementation of such operators on
the space of wavefunctions that are solutions to the diffeomorphism
constraint is pretty hopeless, since the Hamiltonian by definition
will have an action that maps out of that space of functions, since it is not
diffeomorphism invariant. In practice,
this is assured by the presence of the regularization dependent prefactors.

One particular proposal, which appeared in the context of a lattice
regularization of the theory \cite{GaPu96b,FoGaPu}, is to ignore the
prefactors and concentrate on the topological part of the operator. By
doing so, we ensure that the kernel of the operator we are considering
is contained in the kernel of the Hamiltonian. On the other hand the
regularization dependent prefactors can, on rather general grounds, be
chosen to be nonzero (this of course is a delicate issue) in a first
analysis. It turns out that a simple regularization based on lattices
yields to a topological action as that shown in figure 3. In this
action the Hamiltonian only acts at triple intersections $\eta_i$ and
it displaces one of the strands going through the intersection along a
direction determined by one of the other strands back ($\eta_2$) and
forth ($\eta_1)$. The action of the operator consists in summing
similar contributions per each pair of tangents incoming to the
intersection. The contribution is completed by rerouting the subloop
determined by the two strands considered in the operation, as indicated by
the numbers in the figure that keep track of the orientation.
\begin{figure}
\hskip 2cm \psfig{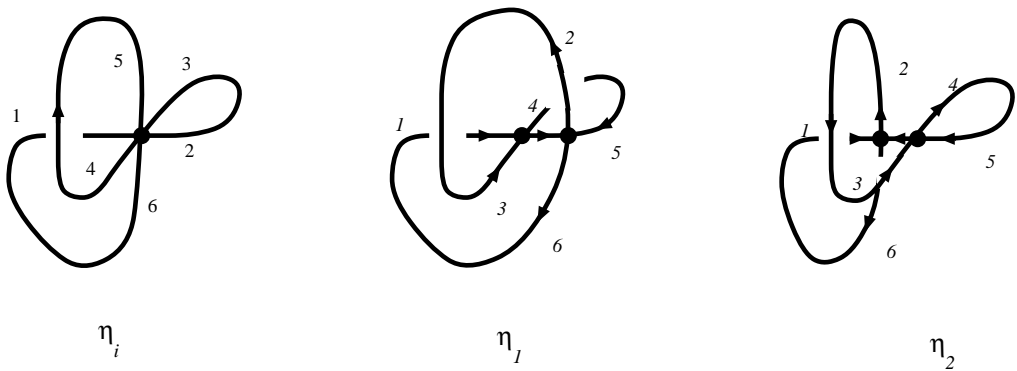}
\caption{The action of the regularized Hamiltonian is only nontrivial at
triple intersections $\eta_i$ and consists of picking a pair of strands at
the intersection and moving one of them back and forth along the other.
The total action is the sum of contributions along all possible pairs of
strands and it also includes a rerouting of one of the two subloops defined
by each pair of strands, as indicated by the numbers that keep track of
orientation.}
\end{figure}

The remarkable property that this regularization has is that the
second \cite{GaPu96b} and third \cite{GaGrPu} coefficient of the
infinite expansion of the Jones polynomial considered in \cite{BrGaPu}
are annihilated by this operator. To prove this, one considers the
individual coefficient and uses its skein relations (the relations
that define it as a knot invariant).

One way of viewing this result \cite{GaPu96b} is that the action of
the Hamiltonian is a skein relation in itself. It is the skein
relation defining the invariant that is the general solution of the
quantum Einstein equations. Notice that this skein relation does not
completely characterize an invariant. This is sensible, since we do
not expect the quantum Einstein equations to have a single
solution. It also sheds new light on how to characterize the infinite
number of degrees of freedom of general relativity in terms of the
topological, discrete notions of knot theory. Another upshot of this
construction is that one need not worry about the algebra of
constraints, since by imposing a skein relation one not only requires
that the identity hold, but all possible rearrangements and multiple
applications of it hold too. This makes sense in the context of
diffeomorphism invariant states, where the Hamiltonian should have
vanishing commutator with itself. Viewing the constraint as a skein
relation therefore allows us a new perspective on the issue of the
algebra of constraints. It is also worwhile remarking that in the
Regge--Ponzano approach to $2+1$ gravity \cite{Rei} there is an
analogous situation: the constraint appears as a relation among graphs
that uniquely determines the invariant that is a (in that case
actually the only) solution.

It might come as a surprise to the reader that both the second and
third coefficients of the infinite expansion of the Jones polynomial
(which by the way are Vassiliev invariants \cite{BaNa}) appear as
solutions. In calculations in terms of extended loops, the third
coefficient was not a solution \cite{Gr}.  The answer to this is that
this regularization is different. The calculations have some common
elements with the extended loop ones but in the end lead to different
results. Recall that the operator we are dealing with here is finite
whereas the extended loop one required a regularization, that in turn
required a counterterm to have the second coefficient be a solution
\cite{DiBaGaGr}. We have checked that other possible candidates (for
instance the square of the second coefficient) are not annihilated by
the operator, which therefore has a nontrivial action. It would be
interesting to see if the operator manages to annihilate all the
Vassiliev invariants stemming from the Jones polynomial. The interest
of these results for quantum gravity is still questionable, since only
the second coefficient satisfies the Mandelstam identities and
therefore can be a candidate to state of the quantum gravitational field.
These results however, highlight how different regularization choices
can lead to different properties of the quantum theory. It should be
emphasized that both in the extended loop case and in the skein
relation calculations, the fact that the knot invariants we are
discussing are annihilated by the Hamiltonian constraint arises as a
consequence of a very complex cancellation, involving in a very strong
way properties of the knot invariants. It is remarkable that one can
find two regularizations that differ in simple details and yet lead to
theories with so distinct properties and of so rich a structure.

\section{Conclusions}

In summary, progess in this area is steady. It is very encouraging to see
physics emerging from these considerations, as the first steps towards 
calculations of the Bekenstein bound show \cite{Ro96}. It is also worthwhile
noticing that the various sets of ideas discussed are entangling themselves
in a nontrivial way, yielding a rather unexpected picture of quantum gravity.
One of the most exciting aspects is that we will soon be able to tie 
physics deeply into the mathematics developed: for instance it is unlikely
that all the regularizations of the Hamiltonian constraint proposed will
lead to the correct Bekenstein bound for black holes. It is also quite 
likely that the mathematical developments that allow to put the theory on 
a more rigorous setting will act as a guideline as to what particular 
choices one has to make to construct the theory. The goal of a final theory
of quantum gravity probably is still afar, but the fact that one is able
to establish a framework that is finding its way towards concrete physical
applications and yet retaining some of the original ideas that motivated
it is a quite encouraging development.

\acknowledgements 

I wish to thank the organizers of ICGC 95 for inviting me to speak,
for travel support and hospitality at Poona.  This work was supported
in part by grants NSF-INT-9406269, NSF-INT-9513843, NSF-PHY-9423950,
by funds of the Pennsylvania State University and its Office for
Minority Faculty Development, and the Eberly Family Research Fund at
Penn State. The author also acknowledges support from the Alfred
P. Sloan Foundation through an Alfred P. Sloan fellowship.

\end{document}